\documentclass[a4paper,11pt]{article}

\usepackage[dvips]{hyperref}
\hypersetup{
colorlinks=true,
linkcolor=magenta,
citecolor=blue,
urlcolor=black,
bookmarksnumbered=true,
bookmarkstype=toc}
\usepackage[margin=1in]{geometry}
\usepackage{amsmath,amsbsy}
\usepackage[adobe-utopia]{mathdesign}
\usepackage[T1]{fontenc}
\usepackage{cite}
\usepackage{fancyhdr}
\usepackage[font=small,labelfont=bf]{caption}
\usepackage{subfigure}
\usepackage{tikz}
\usetikzlibrary{arrows}
\makeatletter

\@addtoreset{equation}{section}
\makeatother
\usepackage{titlesec}
\titleformat{\section}[block]{\filright\bfseries\mathversion{bold}}{\thesection.}{0.5em}{}[\titlerule]
\titleformat{\subsection}[block]{\filright\bfseries\mathversion{bold}}{\thesubsection.}{0.5em}{}

\makeatletter

\@addtoreset{equation}{section}
\makeatother

\title{
\Large\bfseries Recurrence Relations for Finite-Temperature Correlators\\ via AdS$_{2}$/CFT$_{1}$
}
\author{
Satoshi Ohya\footnote{\texttt{E-mail:\,\href{mailto:ohyasato@fjfi.cvut.cz}{ohyasato@fjfi.cvut.cz}}}\\[2ex]
\textit{\small Department of Physics, Faculty of Nuclear Sciences and Physical Engineering}\\
\textit{\small Czech Technical University in Prague}\\
\textit{\small Pohrani\v{c}n\'{i} 1288/1, 40501 D\v{e}\v{c}\'{i}n, Czech Republic}\\
\textit{\small and}\\
\textit{\small Doppler Institute for Mathematical Physics and Applied Mathematics}\\
\textit{\small Czech Technical University in Prague}\\
\textit{\small B\v{r}ehov\'{a} 7, 11519 Prague, Czech Republic}
}
\date{\small (Dated: \today)}

\begin{document}
\maketitle
\thispagestyle{fancy}
\renewcommand{\headrulewidth}{0pt}
\rhead{DI13-019}

\begin{abstract}
This note is aimed at presenting a new algebraic approach to momentum-space correlators in conformal field theory.
As an illustration we present a new Lie-algebraic method to compute frequency-space two-point functions for charged scalar operators of CFT$_{1}$ dual to AdS$_{2}$ black hole with constant background electric field.
Our method is based on the real-time prescription of AdS/CFT correspondence, Euclideanization of AdS$_{2}$ black hole and projective unitary representations of the Lie algebra $\mathfrak{sl}(2,\mathbb{R}) \oplus \mathfrak{sl}(2,\mathbb{R})$.
We derive novel recurrence relations for Euclidean CFT$_{1}$ two-point functions, which are exactly solvable and completely determine the frequency- and charge-dependences of two-point functions.
Wick-rotating back to Lorentzian signature, we obtain retarded and advanced CFT$_{1}$ two-point functions that are consistent with the known results.
\end{abstract}

\newpage
\section{Introduction} \label{sec:1}
Ever since the pioneering work of Polyakov \cite{Polyakov:1970xd}, conformal invariance has been shown to be powerful enough to determine correlation functions in critical systems.
It is indeed a well-known result (see e.g. \cite{Osborn:1993cr}) that, up to overall normalization factors, two- and three-point functions of $d$-dimensional conformal field theory (CFT$_{d}$) are completely determined only through conformal symmetry: Together with translation and rotation invariances, covariance properties under dilatation and special conformal transformations completely fix the possible forms of two- and three-point functions of (quasi-)primary operators in any spacetime dimension $d \geq 1$.
Usually these symmetry constraints work well in coordinate space but tell us little about momentum-space correlators before performing Fourier transform.
Indeed, in spite of its simplicity in coordinate space, three-point functions in momentum space are known to be very complicated.
(For recent studies on conformal constraints in momentum-space three-point functions, see \cite{Coriano:2013jba,Bzowski:2013sza}.)
Since retarded correlators in momentum space, for example, are directly related to physical observables such as spectral density and/or conductivities within the linear response approximation, it would be desirable to understand how directly conformal symmetry restricts the possible forms of momentum-space correlators.
From a practical computational point of view, this is also desirable because Fourier transform of position-space correlators is generally hard.
In this note we would like to present a small example that, up to overall normalization factors, finite-temperature CFT$_{1}$ two-point functions in frequency space can also be completely determined directly through symmetry considerations.
The keys to understanding are real-time holography techniques given by Iqbal and Liu \cite{Iqbal:2008by,Iqbal:2009fd} and novel recurrence relations for frequency-space correlators resulting from $SL(2, \mathbb{R})$ isometry of two-dimensional anti-de Sitter (AdS$_{2}$) spacetime that comes from near-horizon geometries of black holes.

To simplify the problem in this note we will focus on scalar field theory on AdS$_{2}$ black hole with constant background electric field.
Here AdS$_{2}$ black hole means a locally AdS$_{2}$ spacetime with a preferred choice of time \cite{Spradlin:1999bn}.
From a symmetry perspective, as we will see later, AdS$_{2}$ black hole is a locally AdS$_{2}$ spacetime whose time-translation generator generates the one-parameter subgroup $SO(1,1) \subset SL(2, \mathbb{R})$, which, after Wick rotation, becomes the compact rotation group $SO(2)$ and hence leads to quantized Matsubara frequencies conjugate to the imaginary time.
Such AdS$_{2}$ black hole with constant background electric field can be ubiquitous in a sense that near-horizon geometries of any (charged) black holes have an AdS$_{2}$ factor (see for review \cite{Kunduri:2013gce}).
For example, in the near-horizon limit the Reissner-Nordstr\"{o}m black hole has the following AdS$_{2}$ factor together with the constant background $U(1)$ gauge field:
\begin{align}
ds_{\text{AdS}_{2}}^{2}
= 	- \left(\left(\frac{r}{r_{0}}\right)^{2} - 1\right)dt^{2}
	+ \frac{dr^{2}}{(r/r_{0})^{2} - 1}
\quad\text{with}\quad
A
= 	-E(r - r_{0})dt, \label{eq:1.1}
\end{align}
where $r_{0}$ is the horizon radius and $E$ is a constant background electric field.
(In two-dimensional spacetimes electric field is said to be constant if the field strength 2-form is just proportional to the volume element, $F = dA = E\sqrt{|g|}dt \wedge dr$.)
In the following discussions it is convenient to work in a conformal metric by introducing a new spatial coordinate $x$ via
\begin{align}
r
= 	r_{0}\coth(x/r_{0}). \label{eq:1.2}
\end{align}
In the coordinate system $(t, x)$, where the horizon and AdS$_{2}$ boundary are located at $x=\infty$ and $x=0$, respectively, the metric and gauge field become
\begin{align}
ds_{\text{AdS}_{2}}^{2}
= 	\frac{- dt^{2} + dx^{2}}{\sinh^{2}(x/r_{0})}
\quad\text{and}\quad
A
= 	-Er_{0}(\coth(x/r_{0}) - 1)dt. \label{eq:1.3}
\end{align}
Following Ref.~\cite{Spradlin:1999bn} we call this coordinate system $(t, x)$ the Schwarzschild coordinates.
Near-horizon geometries of other charged black holes such as Reissner-Nordstr\"{o}m-AdS black hole (see footnote \ref{footnote:6}) have the same AdS$_{2}$ factor as Eq.~\eqref{eq:1.3} by appropriate change of spatial coordinate (and rescaling of time).
It should be emphasized here that, though the metric \eqref{eq:1.3} can be written into the standard global or Poincar\'{e} metric by spacetime coordinate transformations, such coordinate transformations induce mixings of positive- and negative-frequency modes such that the field theory vacuum in the Schwarzschild coordinates is different from those in the global and Poincar\'{e} coordinates \cite{Spradlin:1999bn}.
Throughout this note we will work in the Schwarzschild coordinates.

The purpose of this note is to present a new Lie-algebraic method to compute real-time finite-temperature two-point functions in frequency space for charged scalar operators of dual CFT$_{1}$ living on the boundary of AdS$_{2}$ black hole \eqref{eq:1.3}.
We emphasize that such CFT$_{1}$ two-point functions are already computed by Faulkner \textit{et al.} \cite{Faulkner:2009wj,Faulkner:2011tm} (see also review \cite{Iqbal:2011ae}) in the context of holographic non-Fermi liquids whose low-energy behaviors are governed by a nontrivial infrared fixed point at which only the time direction admits scale invariance.
Making use of the real-time prescription of AdS/CFT correspondence \cite{Iqbal:2008by,Iqbal:2009fd}, they derived frequency-space two-point functions by solving the bulk field equations explicitly.
Though also based on the same real-time AdS/CFT prescription, our Lie-algebraic method does not require to solve the bulk field equation explicitly.
We believe that our method still has some value and gains insight into the roles of Lie algebra of conformal group and its unitary representations in momentum-space CFT two-point functions.
To discuss our method, let us first recall how to compute frequency-space CFT$_{1}$ two-point functions via real-time prescription of AdS/CFT correspondence.

Let $\Phi(x,t)$ be a complex scalar field of charge $q$ living on the background spacetime \eqref{eq:1.3} and satisfy the Klein-Gordon equation $(\Box_{\text{AdS}_{2}} - m^{2})\Phi = 0$, where the d'Alembertian is given by $\Box_{\text{AdS}_{2}} = \sinh^{2}(x/r_{0})[-(\partial_{t} - iqA_{t})^{2} + \partial_{x}^{2}]$.
In order to get CFT$_{1}$ two-point functions in frequency space we need to find positive-frequency mode solutions whose time-dependence are given by $\Phi(x,t) = \Phi_{\omega}(x)\mathrm{e}^{-i\omega t}$.
In other words, we need to know \textit{simultaneous eigenstates} of the d'Alembertian $\Box_{\text{AdS}_{2}}$ and the time-translation generator $i\partial_{t}$.
For such a simultaneous eigenstate the Klein-Gordon equation reduces to the following one-dimensional Schr\"{o}dinger equation with Eckart potential:\footnote{For Eckart potential see e.g. \cite{Derezinski:2010} and references therein.}
\begin{align}
\left[
- r_{0}^{2}\partial_{x}^{2}
+ \frac{\Delta(\Delta - 1)}{\sinh^{2}(x/r_{0})}
- 2\alpha(\omega r_{0} + \alpha)\coth(x/r_{0})
\right]
\Phi_{\omega}
&= 	\left[
	(\omega r_{0} + \alpha)^{2} + \alpha^{2}
	\right]\Phi_{\omega}, \label{eq:1.4}
\end{align}
where $\alpha := qEr_{0}^{2}$ is a dimensionless parameter, which we hereafter call the coupling constant, and $\Delta$ is the conformal weight of dual CFT$_{1}$ operator given by
\begin{align}
\Delta
&= 	\frac{1}{2} + \sqrt{(mr_{0})^{2} - (m_{\text{BF}}r_{0})^{2}}, \label{eq:1.5}
\end{align}
with $(m_{\text{BF}}r_{0})^{2} := \alpha^{2} - 1/4$ being the Breitenlohner-Freedman bound on AdS$_{2}$ with constant background electric field \cite{Pioline:2005pf}.
Noting that the Eckart potential $V(x) = \Delta(\Delta - 1)/\sinh^{2}(x/r_{0}) - 2\alpha(\omega r_{0} + \alpha)\coth(x/r_{0})$ asymptotes near the boundary as $V(x) = r_{0}^{2}\Delta(\Delta - 1)/x^{2} + O(1/x)$, we see that the right-hand side of Eq.~\eqref{eq:1.4} can be neglected in the leading order $O(1/x^{2})$ and the general solution has the following near-boundary behavior:
\begin{align}
\Phi(x, t)
&\sim
	A_{\Delta}(\omega, \alpha)x^{\Delta}\mathrm{e}^{-i\omega t}
	+ B_{\Delta}(\omega, \alpha)x^{1-\Delta}\mathrm{e}^{-i\omega t}
	\quad\text{as}\quad x\to0, \label{eq:1.6}
\end{align}
where $A_{\Delta}(\omega, \alpha)$ and $B_{\Delta}(\omega, \alpha)$ are integration constants which may depend on $\Delta$, $\omega$ and $\alpha$.
According to the real-time prescription of AdS/CFT correspondence \cite{Iqbal:2008by,Iqbal:2009fd}, the retarded/advanced two-point functions for charged scalar operator of conformal weight $\Delta$ in dual CFT$_{1}$ living on the boundary $x=0$ are given by the ratio
\begin{align}
G^{R/A}_{\Delta}(\omega, \alpha)
&= 	(2\Delta - 1)\frac{A_{\Delta}(\omega, \alpha)}{B_{\Delta}(\omega, \alpha)}, \label{eq:1.7}
\end{align}
where the retarded two-point function $G^{R}_{\Delta}$ corresponds to impose the in-falling boundary condition at the horizon, whereas the advanced two-point function $G^{A}_{\Delta}$ corresponds to impose the out-going boundary condition at the horizon \cite{Son:2002sd}.
The goal of this note is to show that, up to an $\omega$- and $\alpha$-independent normalization factor, the ratio \eqref{eq:1.7} can be completely determined by symmetry without solving the field equation explicitly.
The keys to understanding our method are unitary representations of the Lie algebra $\mathfrak{sl}(2,\mathbb{R})$ of the isometry group $SL(2,\mathbb{R})$ for \textit{Euclidean} AdS$_{2}$ black hole.
Before going into details, let us briefly outline the central idea of the method.

As is well-known, the isometry group for both Lorentzian and Euclidean AdS$_{2}$ is the special linear group $SL(2, \mathbb{R})$.\footnote{To be precise, the isometry group of AdS$_{2}$ is $SL(2,\mathbb{R})/\mathbb{Z}_{2} \cong SO(2,1)$. (Notice that $SL(2,\mathbb{R})$ and $SO(2,1)$ are in two-to-one correspondence.) However, this $\mathbb{Z}_{2}$ identification has little to do with the following Lie-algebraic discussions.}
It is also well-known that $SL(2, \mathbb{R})$ contains three distinct one-parameter subgroups: compact rotation group $SO(2)$, non-compact Euclidean group $E(1)$ and non-compact Lorentz group $SO(1,1)$.
Correspondingly, there exist three distinct classes of AdS$_{2}$ coordinate systems in which time-translation generators generate the one-parameter subgroups $SO(2)$, $E(1)$ and $SO(1,1)$, respectively.
In conformal gauge, such coordinates are turned out to be given, respectively, by global coordinates, Poincar\'e coordinates and Schwarzschild coordinates.
Meanwhile, unitary representations of the Lie algebra $\mathfrak{sl}(2,\mathbb{R})$, which, in the Cartan-Weyl basis, is spanned by the compact $SO(2)$ generator $J_{3}$ and the raising- and lowering-operators $J_{\pm}$, are usually based on simultaneous eigenstates of $J_{3}$ and the quadratic Casimir of the Lie algebra $\mathfrak{sl}(2,\mathbb{R})$.\footnote{Unitary representations of the Lie algebra $\mathfrak{su}(1,1) \cong \mathfrak{sl}(2,\mathbb{R})$ in the basis in which the quadratic Casimir and the non-compact $E(1)$ or $SO(1,1)$ generator become diagonal are studied in \cite{Lindblad:1970}.}
Eigenvalues of the compact generator $J_{3}$ are quantized and raised and lowered by $J_{\pm}$.
The actions of $J_{\pm}$ on the simultaneous eigenstates yield ladder equations with respect to the eigenvalue of $J_{3}$.
In our problem, we need to know simultaneous eigenstates of the d'Alembertian and the time-translation generator.
The d'Alembertian is given by the quadratic Casimir of the Lie algebra $\mathfrak{sl}(2,\mathbb{R})$ of the isometry group $SL(2,\mathbb{R})$ for (both Lorentzian and Euclidean) AdS$_{2}$ black hole.
The time-translation generator, on the other hand, is given by the non-compact $SO(1,1)$ generator in Lorentzian Schwarzschild coordinates; However, it is given by the compact $SO(2)$ generator in Euclidean Schwarzschild coordinates.
Hence, in Euclidean signature, simultaneous eigenstates of the d'Alembertian and the time-translation generator must satisfy ladder equations with respect to the eigenvalue of Wick-rotated time-translation generator (i.e. Matsubara frequency).
These ladder equations turn out to induce some recurrence relations for the integration constants $A_{\Delta}(\omega,\alpha)$ and $B_{\Delta}(\omega,\alpha)$ in Eq.~\eqref{eq:1.6} with respect to the Matsubara frequency.
As we will see below, these recurrence relations are exactly solvable such that the frequency dependence of the ratio \eqref{eq:1.7} is completely determined, even without solving the field equation explicitly.
A remarkable point is that the $\alpha$-dependence of the two-point function \eqref{eq:1.7} is also completely determined in a similar Lie-algebraic way by introducing a \textit{fictitious} extra dimension $S^{1}$ and promoting the Wick-rotated coupling constant to (minus) the conjugate momentum operator $i\partial_{\theta}$ of coordinate $\theta \in S^{1}$.

The rest of the note is organized as follows:
In section \ref{sec:2} we review $SL(2,\mathbb{R})$ isometry of Euclidean AdS$_{2}$ black hole in the presence of constant background electric field.
We then introduce a spectrum generating algebra $\mathfrak{sl}(2,\mathbb{R})_{A} \oplus \mathfrak{sl}(2,\mathbb{R})_{B}$ on $\text{AdS}_{2} \times S^{1}$ in section \ref{sec:3} and its asymptotic near-boundary algebra $\mathfrak{sl}(2,\mathbb{R})_{A}^{0} \oplus \mathfrak{sl}(2,\mathbb{R})_{B}^{0}$ realized at the AdS$_{2}$ boundary $x = 0$ in section \ref{sec:4}, where $\mathfrak{sl}(2,\mathbb{R})_{A}$ is the Lie algebra $\mathfrak{sl}(2,\mathbb{R})$ whose $SO(2)$ generator is the time-translation generator whereas $\mathfrak{sl}(2,\mathbb{R})_{B}$ is the Lie algebra $\mathfrak{sl}(2,\mathbb{R})$ whose $SO(2)$ generator is the coupling constant operator $i\partial_{\theta}$.
By making use of the boundary spectrum generating algebra $\mathfrak{sl}(2,\mathbb{R})_{A}^{0} \oplus \mathfrak{sl}(2,\mathbb{R})_{B}^{0}$, in section \ref{sec:5} we derive recurrence relations for Euclidean two-point functions, which can be solved exactly up to an overall frequency- and coupling-constant-independent normalization factor.
Wick-rotating back to Lorentzian signature, we obtain the retarded/advanced two-point functions in frequency space, which are consistent with \cite{Faulkner:2009wj,Faulkner:2011tm}.
Section \ref{sec:6} is devoted to conclusions and discussions.

Below we will work in the units in which $r_{0} = 1$ unless otherwise stated.

\section{\texorpdfstring{$SL(2,\mathbb{R})$}{SL(2,R)} isometry of Euclidean \texorpdfstring{AdS$_{2}$}{AdS2} black hole with background electric field} \label{sec:2}
In this section we derive generators of $SL(2,\mathbb{R})$ symmetry transformations for charged scalar fields residing on Euclidean AdS$_{2}$ black hole with background $U(1)$ gauge field.

To begin with, let us consider Euclideanization of AdS$_{2}$ black hole \eqref{eq:1.3} by taking the analytic continuations
\begin{align}
t_{E} = it
\quad\text{and}\quad
E_{E} = -iE, \label{eq:2.1}
\end{align}
where $E_{E}$ is the Wick-rotated electric field.
The metric and gauge field in Euclidean signature are therefore (in the units $r_{0} = 1$)
\begin{align}
ds_{\text{EAdS}_{2}}^{2}
= 	\frac{dt_{E}^{2} + dx^{2}}{\sinh^{2}x}
\quad\text{and}\quad
A
= 	-E_{E}(\coth x - 1)dt_{E}. \label{eq:2.2}
\end{align}
It is straightforward to show that the line element is invariant under the $SL(2,\mathbb{R})$ group action
\begin{align}
\begin{pmatrix}
a 	& b \\
c 	& d
\end{pmatrix}:
z
\mapsto z^{\prime}
= 	2\arctan
	\left(\frac{a\tan(z/2) + b}{c\tan(z/2) + d}\right), \label{eq:2.3}
\end{align}
where $z = t_{E} + ix$ and $a,b,c,d \in \mathbb{R}$ with $ad - bc = 1$.
Let us next consider the Lie algebra $\mathfrak{sl}(2,\mathbb{R})$ of the isometry group $SL(2,\mathbb{R})$, which consists of traceless $2 \times 2$ real matrices.
As a basis we choose the following matrices:
\begin{align}
iJ_{1}
= 	\frac{1}{2}
	\begin{pmatrix}
	0 	& 1 \\
	1 	& 0
	\end{pmatrix}, \quad
iJ_{2}
= 	\frac{1}{2}
	\begin{pmatrix}
	1 	& 0 \\
	0 	& -1
	\end{pmatrix}, \quad
iJ_{3}
= 	\frac{1}{2}
	\begin{pmatrix}
	0 	& 1 \\
	-1 	& 0
	\end{pmatrix}, \label{eq:2.4}
\end{align}
which satisfies the commutation relations of the Lie algebra $\mathfrak{sl}(2,\mathbb{R}) \cong \mathfrak{so}(2,1)$
\begin{align}
[J_{1}, J_{2}] = iJ_{3}, \quad
[J_{2}, J_{3}] = -iJ_{1}, \quad
[J_{3}, J_{1}] = -iJ_{2}. \label{eq:2.5}
\end{align}
These matrices generate the following one-parameter subgroups:
\begin{subequations}
\begin{align}
\exp(i\epsilon J_{1})
&= 	\begin{pmatrix}
	\cosh\frac{\epsilon}{2} 	& \sinh\frac{\epsilon}{2} \\
	\sinh\frac{\epsilon}{2} 	& \cosh\frac{\epsilon}{2}
	\end{pmatrix} \in SO(1,1), \label{eq:2.6a}\\
\exp(i\epsilon J_{2})
&= 	\begin{pmatrix}
	\mathrm{e}^{\epsilon/2} 	& 0 \\
	0 					& \mathrm{e}^{-\epsilon/2}
	\end{pmatrix} \in SO(1,1), \label{eq:2.6b}\\
\exp(i\epsilon J_{3})
&= 	\begin{pmatrix}
	\cos\frac{\epsilon}{2} 	& \sin\frac{\epsilon}{2} \\
	-\sin\frac{\epsilon}{2} 		& \cos\frac{\epsilon}{2}
	\end{pmatrix} \in SO(2), \label{eq:2.6c}
\end{align}
\end{subequations}
which, up to a linear order in $\epsilon$, turn out to induce the coordinate transformations
$\exp(i\epsilon J_{1}):
z
\mapsto z^{\prime}
= 	z + \epsilon\cos z + O(\epsilon^{2})$,
$\exp(i\epsilon J_{2}):
z
\mapsto z^{\prime}
= 	z + \epsilon\sin z + O(\epsilon^{2})$ and
$\exp(i\epsilon J_{3}):
z
\mapsto z^{\prime}
= 	z + \epsilon$,
respectively.
In terms of the original coordinates $x^{\mu} = (t_{E}, x)$ these infinitesimal transformations read $\exp(i\epsilon J_{a}): x^{\mu} \mapsto x^{\prime\mu} = x^{\mu} + \delta_{a}x^{\mu}$ ($a=1,2,3$), where
\begin{subequations}
\begin{align}
\exp(i\epsilon J_{1}):
&\begin{cases}
\delta_{1}t_{E} = \epsilon\cos t_{E}\cosh x + O(\epsilon^{2}), \\
\delta_{1}x = -\epsilon\sin t_{E}\sinh x + O(\epsilon^{2}),
\end{cases} \label{eq:2.7a}\\
\exp(i\epsilon J_{2}):
&\begin{cases}
\delta_{2}t_{E} = \epsilon\sin t_{E}\cosh x + O(\epsilon^{2}), \\
\delta_{2}x = \epsilon\cos t_{E}\sinh x + O(\epsilon^{2}),
\end{cases} \label{eq:2.7b}\\
\exp(i\epsilon J_{3}):
&\begin{cases}
\delta_{3}t_{E} = \epsilon, \\
\delta_{3}x = 0.
\end{cases} \label{eq:2.7c}
\end{align}
\end{subequations}
Notice that the action of compact rotation group $SO(2)$ gives rise to the time-translation \eqref{eq:2.7c}.
These coordinate transformations leave the line element unchanged.
However, the gauge field does not remain unchanged under the $SL(2,\mathbb{R})$ group action.
As usual, vector fields transform under the coordinate transformations $x^{\mu} \mapsto x^{\prime\mu} = x^{\mu} + \delta x^{\mu}$ as $A_{\mu}(x) \mapsto A_{\mu}^{\prime}(x^{\prime}) = \frac{\partial x^{\nu}}{\partial x^{\prime\mu}}A_{\nu}(x)$.
Corresponding intrinsic field variations are $\delta A_{\mu}(x) = A_{\mu}^{\prime}(x) - A_{\mu}(x) = - \delta x^{\nu}\partial_{\nu}A_{\mu}(x) - A_{\nu}(x)\partial_{\mu}\delta x^{\nu} + O(\delta x)^{2}$, which turn out to become the following $U(1)$ gauge transformations:
\begin{align}
\delta_{a}A_{\mu}(x)
&= 	\epsilon\partial_{\mu}\Lambda_{a}(x) + O(\epsilon^{2}), \label{eq:2.8}
\end{align}
where $\Lambda_{a}$ ($a=1,2,3$) are scalar functions given by
\begin{subequations}
\begin{align}
\Lambda_{1}
&= 	-E_{E}\cos t_{E}(\cosh x - \sinh x), \label{eq:2.9a}\\
\Lambda_{2}
&= 	-E_{E}\sin t_{E}(\cosh x - \sinh x), \label{eq:2.9b}\\
\Lambda_{3}
&= 	-E_{E}. \label{eq:2.9c}
\end{align}
\end{subequations}
We emphasize that, though the gauge field is invariant under the time-translation \eqref{eq:2.7c}, we here choose $\Lambda_{3}$ to be nonzero constant for later convenience.
Let us turn to the transformation law of scalar field $\Phi$.
For the case of neutral scalar field, $\Phi$ transforms under the coordinate transformations $x^{\mu} \mapsto x^{\prime\mu} = x^{\mu} + \delta x^{\mu}$ as $\Phi(x) \mapsto \Phi^{\prime}(x^{\prime}) = \Phi(x)$ such that the intrinsic field variation is given by $\delta\Phi(x) = \Phi^{\prime}(x) - \Phi(x) = -\delta x^{\mu}\partial_{\mu}\Phi(x) + O(\delta x)^{2}$.
For the case of charged scalar field, on the other hand, transformation law of $\Phi$ must be accompanied with the $U(1)$ gauge transformations in order to compensate for the gauge transformation \eqref{eq:2.8}.
Hence the charged scalar field $\Phi$ of charge $q$, which couples to the background gauge field via the covariant derivative $D_{\mu} = \partial_{\mu} - iqA_{\mu}(x)$, must transform as follows:
\begin{align}
\delta_{a}\Phi(x)
&= 	- \delta_{a}x^{\mu}\partial_{\mu}\Phi(x)
	+ i\epsilon q\Lambda_{a}(x)\Phi(x)
	+ O(\epsilon^{2}) \nonumber\\
&=: 	i\epsilon J_{a}\Phi(x) + O(\epsilon^{2}), \label{eq:2.10}
\end{align}
where $J_{a}$ ($a=1,2,3$) are the first-order differential operators given by $i\epsilon J_{a} = -\delta_{a}x^{\mu}\partial_{\mu} + i\epsilon q\Lambda_{a}$.\footnote{Here are the explicit expressions:
\begin{align}
iJ_{1}
&= 	\cos t_{E}\cosh x(-\partial_{t_{E}} - i\alpha_{E})
	+\sin t_{E}\sinh x\partial_{x}
	+i\alpha_{E}\cos t_{E}\sinh x, \nonumber\\
iJ_{2}
&= 	\sin t_{E}\cosh x(-\partial_{t_{E}} - i\alpha_{E})
	-\cos t_{E}\sinh x\partial_{x}
	+i\alpha_{E}\sin t_{E}\sinh x, \nonumber\\
iJ_{3}
&= 	-\partial_{t_{E}} - i\alpha_{E}. \nonumber
\end{align}}
For the following discussions it is convenient to perform the field redefinition $\Phi \mapsto \Tilde{\Phi} = \mathrm{e}^{i\alpha_{E}t_{E}}\Phi$, which induces the similarity transformations $J_{a} \mapsto \Tilde{J}_{a} = \mathrm{e}^{i\alpha_{E}t_{E}}J_{a}\mathrm{e}^{-i\alpha_{E}t_{E}}$ that effectively shift $\partial_{t_{E}}$ to $\partial_{t_{E}} - i\alpha_{E}$.
Then in the Cartan-Weyl basis $\{\Tilde{J}_{3}, \Tilde{J}_{\pm} = -\Tilde{J}_{1} \pm i\Tilde{J}_{2}\}$ the differential operators take the following forms:
\begin{subequations}
\begin{align}
\Tilde{J}_{3}
&=
 	i\partial_{t_{E}}, \label{eq:2.11a}\\
\Tilde{J}_{\pm}
&=
 	\mathrm{e}^{\mp it_{E}}\sinh x
	\left[
	\mp\partial_{x} - \coth x(i\partial_{t_{E}}) - \alpha_{E}
	\right], \label{eq:2.11b}
\end{align}
\end{subequations}
where $\alpha_{E} := qE_{E} (= qE_{E}r_{0}^{2})$ is a dimensionless coupling constant in Euclidean signature.
These differential operators are generators of $SL(2,\mathbb{R})$ symmetry transformations and satisfy the following commutation relations of the Lie algebra $\mathfrak{sl}(2,\mathbb{R}) \cong \mathfrak{so}(2,1)$:
\begin{align}
[\Tilde{J}_{3}, \Tilde{J}_{\pm}] = \pm \Tilde{J}_{\pm}, \quad
[\Tilde{J}_{+}, \Tilde{J}_{-}] = - 2\Tilde{J}_{3}. \label{eq:2.12}
\end{align}
The quadratic Casimir $\Tilde{C}_{2}$ of the Lie algebra $\mathfrak{sl}(2,\mathbb{R})$ yields Euclidean d'Alembertian (i.e. Laplace operator) on Euclidean AdS$_{2}$ black hole with constant background electric field
\begin{align}
\Tilde{C}_{2}(\mathfrak{sl}(2,\mathbb{R}))
&= 	- \Tilde{J}_{1}^{2} - \Tilde{J}_{2}^{2} + \Tilde{J}_{3}^{2}
= 	\Tilde{J}_{3}(\Tilde{J}_{3} \pm 1) - \Tilde{J}_{\mp}\Tilde{J}_{\pm} \nonumber\\
&= 	\sinh^{2}x
	\left[
	\partial_{x}^{2} + \partial_{t_{E}}^{2} - \alpha_{E}^{2}
	- 2\alpha_{E}\coth x(i\partial_{t_{E}})
	\right], \label{eq:2.13}
\end{align}
which commutes with all the generators, $[\Tilde{C}_{2}, \Tilde{J}_{a}] = 0$ ($a = 3, +, -$), as it should.
Let $\Tilde{\Phi}_{\Delta, \omega_{E}}$ be a simultaneous eigenstate of $\Tilde{C}_{2}$ and $\Tilde{J}_{3}$ with eigenvalues $\Delta(\Delta-1)$ and $\omega_{E}$, respectively.
Then the eigenvalue equation $\Tilde{C}_{2}\Tilde{\Phi}_{\Delta, \omega_{E}} = \Delta(\Delta-1)\Tilde{\Phi}_{\Delta, \omega_{E}}$ reduces to the following negative-energy bound state problem of Schr\"{o}dinger equation with Eckart potential:
\begin{align}
\left(
-\partial_{x}^{2} + \frac{\Delta(\Delta-1)}{\sinh^{2}x} + 2\omega_{E}\alpha_{E}\coth x
\right)
\Tilde{\Phi}_{\Delta, \omega_{E}}
&= 	- \left(
	\omega_{E}^{2} + \alpha_{E}^{2}
	\right)
	\Tilde{\Phi}_{\Delta, \omega_{E}}, \label{eq:2.14}
\end{align}
which is nothing but the Euclidean version of Eq.~\eqref{eq:1.4}.
Indeed, by restoring $r_{0}$ via the replacement $\omega_{E} \to \omega_{E}r_{0}$ we recover the original equation \eqref{eq:1.4} via the following relations:
\begin{align}
\omega_{E}r_{0} = -i(\omega r_{0} + \alpha)
\quad\text{and}\quad
\alpha_{E} = -i\alpha. \label{eq:2.15}
\end{align}

Let us close this section with a remark on the boundary condition for the scalar field $\Tilde{\Phi}$.
As in finite-temperature field theory we impose the periodic boundary condition for the original scalar field $\Phi$ with respect to the imaginary time, $\Phi(x, t_{E} + 2\pi) = \Phi(x, t_{E})$.
Then the redefined field $\Tilde{\Phi}(x, t_{E}) = \mathrm{e}^{i\alpha_{E}t_{E}}\Phi(x, t_{E})$ satisfies the following twisted boundary condition:
\begin{align}
\Tilde{\Phi}(x, t_{E} + 2\pi)
&= 	\mathrm{e}^{i2\pi \alpha_{E}}\Tilde{\Phi}(x, t_{E}). \label{eq:2.16}
\end{align}
As we will see in the next section this twisted boundary condition leads to projective unitary representations of the Lie algebra $\mathfrak{sl}(2, \mathbb{R})$.

\section{Spectrum generating algebra \texorpdfstring{$\mathfrak{sl}(2,\mathbb{R})_{A} \oplus \mathfrak{sl}(2,\mathbb{R})_{B}$}{sl(2,R) + sl(2,R)}} \label{sec:3}
As mentioned in the introduction, we shall introduce a fictitious extra dimension $S^{1}$ in which eigenvalues of the conjugate momentum of coordinate of $S^{1}$ give an infinite set of Wick-rotated coupling constants $\{\alpha_{E}\}$.
The original problem on Euclidean AdS$_{2}$ black hole then corresponds to a problem on a fixed $\alpha_{E}$ sector of the Hilbert space on $\text{AdS}_{2} \times S^{1}$.
The advantage of this trick is that we can introduce \textit{two} orthogonal $SL(2,\mathbb{R})$-Lie algebras, $\mathfrak{sl}(2,\mathbb{R}) \oplus \mathfrak{sl}(2,\mathbb{R})$, one is essentially the Lie algebra $\mathfrak{sl}(2,\mathbb{R})$ introduced in the previous section and the other is the Lie algebra $\mathfrak{sl}(2,\mathbb{R})$ whose $SO(2)$ generators is the conjugate momentum operator on $S^{1}$.
In subsequent sections we will see that, by using ladder equations of the Lie algebra $\mathfrak{sl}(2,\mathbb{R}) \oplus \mathfrak{sl}(2,\mathbb{R})$, we can obtain recurrence relations for Euclidean two-point functions not only with respect to the Matsubara frequency but also with respect to the Wick-rotated coupling constant, which enable us to determine frequency- and coupling-constant-dependences of two-point functions completely.
Some readers will notice that these recurrence relations have essentially the same structure as those for quantum mechanical S-matrix (i.e. reflection and transmission amplitudes) with dynamical symmetry \cite{Frank:1984,Alhassid:1984uy}.
Indeed, much of the present work is inspired by Lie-algebraic approach to S-matrix in quantum mechanics developed by Frank and Wolf \cite{Frank:1984} and Alhassid and his collaborators \cite{Alhassid:1984uy}.
A crucial difference is that, in scattering problem one needs to know asymptotic Lie algebra at spatial infinity where all the interactions are switched off, whereas in AdS/CFT one needs to know asymptotic Lie algebra at AdS boundary where potential terms in field equations do not vanish.
After introducing the spectrum generating algebra $\mathfrak{sl}(2,\mathbb{R}) \oplus \mathfrak{sl}(2,\mathbb{R})$ we study its unitary representations realized in the scalar field theory on $\text{AdS}_{2} \times S^{1}$ and derive the spectrum in a purely algebraic fashion.

\subsection{\texorpdfstring{$\text{AdS}_{2} \to \text{AdS}_{2} \times S^{1}$}{AdS2 to AdS2 x S1}}
Now, let us introduce a fictitious spatial compact dimension $S^{1}$ and extend the spacetime as follows:
\begin{align}
\text{AdS$_{2}$} \to \text{AdS$_{2}$} \times S^{1}. \label{eq:3.1}
\end{align}
Let $\theta \in [0, 2\pi)$ be the coordinate of $S^{1}$.
The heart of our Lie-algebraic method is to promote the coupling constant $\alpha_{E}$ to (minus) the conjugate momentum of coordinate $\theta \in S^{1}$:
\begin{align}
\alpha_{E} \to i\partial_{\theta}. \label{eq:3.2}
\end{align}
We now introduce the following differential operators on $\text{AdS}_{2} \times S^{1}$ by replacing $\alpha_{E}$'s in \eqref{eq:2.11a} and \eqref{eq:2.11b} with the coupling constant operator $i\partial_{\theta}$:
\begin{subequations}
\begin{align}
A_{3}
&:= 	i\partial_{t_{E}}, \label{eq:3.3a}\\
A_{\pm}
&:= 	\mathrm{e}^{\mp it_{E}}\sinh x
	\left[
	\mp\partial_{x} - \coth x(i\partial_{t_{E}}) - i\partial_{\theta}
	\right], \label{eq:3.3b}
\end{align}
\end{subequations}
which still span the Lie algebra $\mathfrak{sl}(2,\mathbb{R})$.
Henceforth we denote this Lie algebra by $\mathfrak{sl}(2,\mathbb{R})_{A}$.
We further introduce another Lie algebra $\mathfrak{sl}(2,\mathbb{R})_{B}$ which is orthogonal to $\mathfrak{sl}(2,\mathbb{R})_{A}$ and spanned by the following differential operators:
\begin{subequations}
\begin{align}
B_{3}
&:= 	i\partial_{\theta}, \label{eq:3.4a}\\
B_{\pm}
&:= 	\mathrm{e}^{\mp i\theta}\sinh x
	\left[
	\mp\partial_{x} - \coth x(i\partial_{\theta}) - i\partial_{t_{E}}
	\right]. \label{eq:3.4b}
\end{align}
\end{subequations}
Notice that $A_{+}$ ($B_{+}$) and $A_{-}$ ($B_{-}$) are hermitian conjugate with each other with respect to the integration measure $\sqrt{|g|}d^{2}xd\theta = dt_{E}dxd\theta/\sinh^{2}x$.
It is straightforward to verify that these differential operators indeed satisfy the commutation relations of the Lie algebra $\mathfrak{sl}(2,\mathbb{R})_{A} \oplus \mathfrak{sl}(2,\mathbb{R})_{B} (\cong \mathfrak{so}(2,2))$
\begin{subequations}
\begin{alignat}{3}
&[A_{3}, A_{\pm}] = \pm A_{\pm},&\quad
&[A_{+}, A_{-}] = -2A_{3},& \label{eq:3.5a}\\
&[B_{3}, B_{\pm}] = \pm B_{\pm},&\quad
&[B_{+}, B_{-}] = -2B_{3},& \label{eq:3.5b}
\end{alignat}
\end{subequations}
with other commutators vanishing, $[A_{a}, B_{b}] = 0$ ($a,b=3,+,-$).
The quadratic Casimirs of $\mathfrak{sl}(2,\mathbb{R})_{A}$ and $\mathfrak{sl}(2,\mathbb{R})_{B}$ are $C_{2}(\mathfrak{sl}(2,\mathbb{R})_{A}) = A_{3}(A_{3} \pm 1) - A_{\mp}A_{\pm}$ and $C_{2}(\mathfrak{sl}(2,\mathbb{R})_{B}) = B_{3}(B_{3} \pm 1) - B_{\mp}B_{\pm}$, which turn out to coincide and are given by
\begin{align}
C_{2}(\mathfrak{sl}(2,\mathbb{R})_{A})
= 	C_{2}(\mathfrak{sl}(2,\mathbb{R})_{B})
= 	\sinh^{2}x
	\left(
	\partial_{x}^{2}
	- A_{3}^{2} - B_{3}^{2}
	- 2\coth xA_{3}B_{3}
	\right)
=: 	C_{2}. \label{eq:3.6}
\end{align}
We are interested in simultaneous eigenstates of the time-translation operator, the coupling constant operator and the Laplace operator on $\text{AdS}_{2} \times S^{1}$.
Let $|\Delta, \omega_{E}, \alpha_{E}\rangle$ be such a normalized simultaneous eigenstate of $C_{2}$, $A_{3}$ and $B_{3}$ that satisfies the following eigenvalue equations:
\begin{subequations}
\begin{align}
C_{2}|\Delta,\omega_{E},\alpha_{E}\rangle
&= 	\Delta(\Delta-1)|\Delta,\omega_{E},\alpha_{E}\rangle, \label{eq:3.7a}\\
A_{3}|\Delta,\omega_{E},\alpha_{E}\rangle
&= 	\omega_{E}|\Delta,\omega_{E},\alpha_{E}\rangle, \label{eq:3.7b}\\
B_{3}|\Delta,\omega_{E},\alpha_{E}\rangle
&= 	\alpha_{E}|\Delta,\omega_{E},\alpha_{E}\rangle. \label{eq:3.7c}
\end{align}
\end{subequations}
The commutation relations \eqref{eq:3.5a} and \eqref{eq:3.5b} imply that $A_{\pm}$ and $B_{\pm}$ raise and lower the eigenvalues $\omega_{E}$ and $\alpha_{E}$ by unit steps, $A_{\pm}|\Delta, \omega_{E}, \alpha_{E}\rangle \propto |\Delta, \omega_{E} \pm 1, \alpha_{E}\rangle$ and $B_{\pm}|\Delta, \omega_{E}, \alpha_{E}\rangle \propto |\Delta, \omega_{E}, \alpha_{E} \pm 1\rangle$.
The proportional coefficients can be fixed by computing the squared norms $\|A_{\pm}|\Delta, \omega_{E}, \alpha_{E}\rangle\|^{2}$ and $\|B_{\pm}|\Delta, \omega_{E}, \alpha_{E}\rangle\|^{2}$.\footnote{
$\|A_{\pm}|\Delta, \omega_{E}, \alpha_{E}\rangle\|^{2}
= 	\langle \Delta, \omega_{E}, \alpha_{E}|A_{\mp}A_{\pm}|\Delta, \omega_{E}, \alpha_{E}\rangle
= 	\langle \Delta, \omega_{E}, \alpha_{E}|
	\left(A_{3}(A_{3} \pm 1) - C_{2}\right)
	|\Delta, \omega_{E}, \alpha_{E}\rangle
= 	\omega_{E}(\omega_{E} \pm 1) - \Delta(\Delta - 1)
= 	(\omega_{E} \pm \Delta)(\omega_{E} \pm 1 \mp \Delta).$
Similarly, $\|B_{\pm}|\Delta, \omega_{E}, \alpha_{E}\rangle\|^{2} = (\alpha_{E} \pm \Delta)(\alpha_{E} \pm 1 \mp \Delta)$.}
The results are
\begin{subequations}
\begin{align}
A_{\pm}|\Delta, \omega_{E}, \alpha_{E}\rangle
&= 	\sqrt{(\omega_{E} \pm \Delta)(\omega_{E} \pm 1 \mp \Delta)}
	|\Delta, \omega_{E} \pm 1, \alpha_{E}\rangle, \label{eq:3.8a}\\
B_{\pm}|\Delta, \omega_{E}, \alpha_{E}\rangle
&= 	\sqrt{(\alpha_{E} \pm \Delta)(\alpha_{E} \pm 1 \mp \Delta)}
	|\Delta, \omega_{E}, \alpha_{E} \pm 1\rangle. \label{eq:3.8b}
\end{align}
\end{subequations}
As we will see in the subsequent sections, these ladder equations induce recurrence relations for the integration constants $A_{\Delta}(\omega, \alpha)$ and $B_{\Delta}(\omega, \alpha)$ in \eqref{eq:1.6} with respect to the Matsubara frequency $\omega_{E}$ and the Wick-rotated coupling constant $\alpha_{E}$.
Before discussing these recurrence relations, let us study unitary representations of $\mathfrak{sl}(2,\mathbb{R})_{A} \oplus \mathfrak{sl}(2,\mathbb{R})_{B}$ realized in the scalar field theory on Euclidean AdS$_{2}$ black hole with background electric field.

\subsection{Spectrum and projective unitary representations}
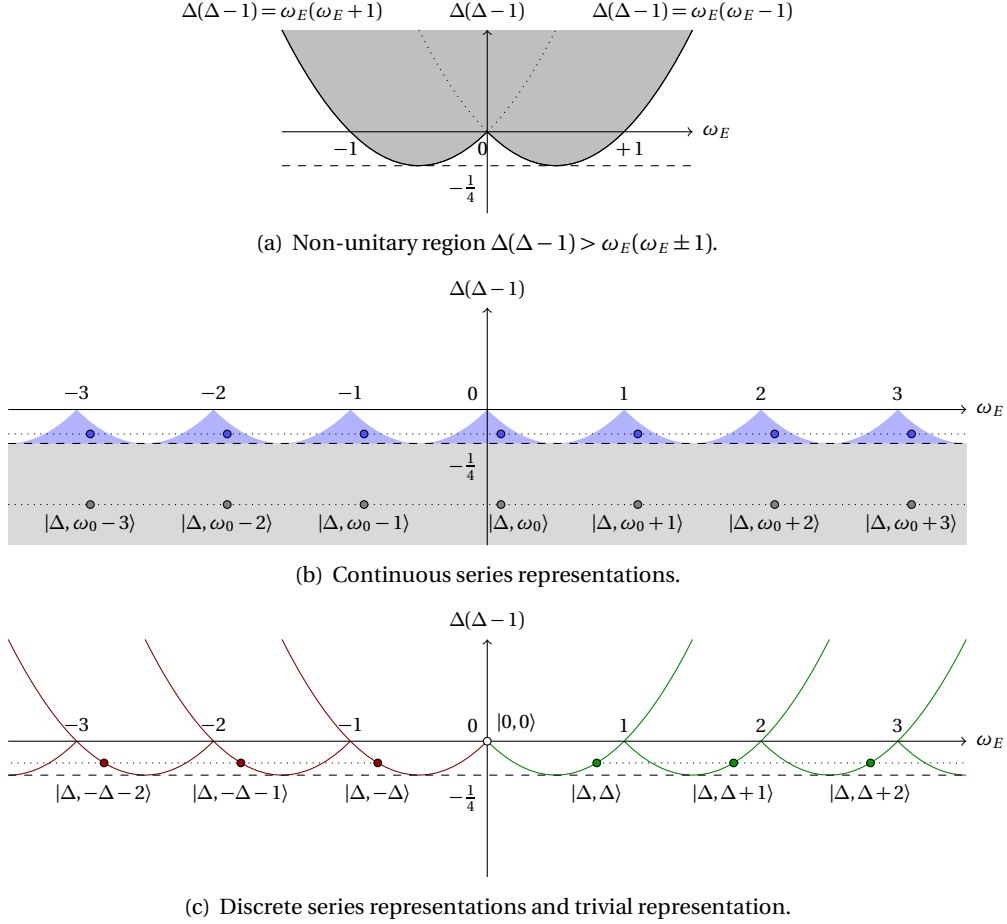
\begin{figure}[t]
\centering
\subfigure[Non-unitary region $\Delta(\Delta-1)>\omega_{E}(\omega_{E}\pm1)$.]{
\begin{tikzpicture} [scale=1.8]
\tikzstyle{every node}=[font=\scriptsize]
\def\XMAX{1.5}
\def\XMIN{-1.5}
\def\YMAX{0.75}
\def\YMIN{-0.6}
\filldraw[fill=gray!50] plot[domain=-1.5:0](\x,{(\x+0.5)*(\x+0.5)-0.25}) -- plot[domain=0:1.5](\x,{(\x-0.5)*(\x-0.5)-0.25});
\draw plot[domain=0:-1.5](\x,{(\x+0.5)*(\x+0.5)-0.25}) node[above]{$\Delta(\Delta-1) = \omega_{E}(\omega_{E}+1)$};
\draw[dotted] plot[domain=0:0.5](\x,{(\x+0.5)*(\x+0.5)-0.25});
\draw plot[domain=0:1.5](\x,{(\x-0.5)*(\x-0.5)-0.25}) node[above]{$\Delta(\Delta-1) = \omega_{E}(\omega_{E}-1)$};
\draw[dotted] plot[domain=-0.5:0](\x,{(\x-0.5)*(\x-0.5)-0.25});
\draw[->] (\XMIN,0) -- (\XMAX,0) node[right]{$\omega_{E}$};
\draw[->] (0,\YMIN) -- (0,\YMAX) node[above]{$\Delta(\Delta-1)$};
\draw[dashed] (\XMIN,-0.25) -- (\XMAX,-0.25);
\draw (0,-0.25) node[below left]{$-\tfrac{1}{4}$};
\draw (0,0) node[below]{$0~~$};
\draw (-1,0) node[below]{$-1~~$};
\draw (1,0) node[below]{$~~+1$};
\end{tikzpicture}
\label{fig:1a}}
\subfigure[Continuous series representations.]{
\begin{tikzpicture} [scale=1.8]
\tikzstyle{every node}=[font=\scriptsize]
\def\XMAX{3.5}
\def\XMIN{-3.5}
\def\YMAX{0.75}
\def\YMIN{-1}
\fill[fill=gray!30] (\XMIN,\YMIN) -- (\XMAX,\YMIN) -- (\XMAX,-0.25) -- (\XMIN,-0.25);
\draw[dotted] (\XMIN,-0.7) -- (\XMAX,-0.7);
\foreach \y in {-3,-2,-1,0,1,2,3}
\filldraw[fill=gray] (\y + 0.1,-0.7) circle (0.8pt);
\foreach \y in {1,2,3}
\draw (\y+0.1,-0.7) node[below]{$|\Delta,\omega_{0}+\y\rangle$} (-\y+0.1,-0.7) node[below]{$|\Delta,\omega_{0}-\y\rangle$};
\draw (0.1,-0.7) node[below]{$~~~~~~~~|\Delta,\omega_{0}\rangle$};
\foreach \y in {-3.5,-2.5,-1.5,-0.5,0.5,1.5,2.5}
\fill[fill=blue!30]
(\y,-0.25) -- plot[domain=\y+1:\y+0.5](\x,{(\x-\y-1)*(\x-\y-1)-0.25}) -- plot[domain=\y+0.5:\y](\x,{(\x-\y)*(\x-\y)-0.25});
\draw[dotted] (\XMIN,-0.18) -- (\XMAX,-0.18);
\foreach \y in {-3,-2,-1,0,1,2,3}
\filldraw[fill=blue!70] (\y + 0.1,-0.18) circle (0.8pt);
\draw[->] (\XMIN,0) node[left]{\phantom{$\omega_{E}$}} -- (\XMAX,0) node[right]{$\omega_{E}$};
\draw[->] (0,\YMIN) -- (0,\YMAX) node[above]{$\Delta(\Delta-1)$};
\draw (0,0) node[above left]{$0$};
\foreach \y in {-3,-2,-1,1,2,3}
\draw (\y,0) node[above]{$\y$};
\draw (0,-0.25) node[below left]{$-\tfrac{1}{4}$};
\draw[dashed] (\XMIN,-0.25) -- (\XMAX,-0.25);
\end{tikzpicture}
\label{fig:1b}}
\subfigure[Discrete series representations and trivial representation.]{
\begin{tikzpicture} [scale=1.8]
\tikzstyle{every node}=[font=\scriptsize]
\def\M0{0.8}
\def\XMAX{3.5}
\def\XMIN{-3.5}
\def\YMAX{0.75}
\def\YMIN{-1}
\foreach \y in {0.5,1.5,2.5}
\draw[green!50!black] plot[domain=\y-0.5:\y+1](\x,{(\x-\y)*(\x-\y)-0.25});
\draw[green!50!black] plot[domain=3:3.5](\x,{(\x-3.5)*(\x-3.5)-0.25});
\draw[dotted] (\M0,\M0*\M0-\M0) -- (\XMAX,\M0*\M0-\M0);
\foreach \y in {\M0,\M0+1,\M0+2} \filldraw[fill=green!60!black] (\y,\M0*\M0-\M0) circle (0.8pt);
\foreach \y in {1,2} \draw (\M0+\y,-0.25) node[below]{$|\Delta,\Delta+\y\rangle$};
\draw (\M0,-0.25) node[below]{$|\Delta,\Delta\rangle$};
\foreach \y in {-2.5,-1.5,-0.5}
\draw[red!50!black] plot[domain=\y+0.5:\y-1](\x,{(\x-\y)*(\x-\y)-0.25});
\draw[red!50!black] plot[domain=-3:-3.5](\x,{(\x+3.5)*(\x+3.5)-0.25});
\draw[dotted] (-\M0,\M0*\M0-\M0) -- (\XMIN,\M0*\M0-\M0);
\foreach \y in {-\M0,-\M0-1,-\M0-2} \filldraw[fill=red!60!black] (\y,\M0*\M0-\M0) circle (0.8pt);
\foreach \y in {1,2} \draw (-\M0-\y,-0.25) node[below]{$|\Delta,-\Delta-\y\rangle$};
\draw (-\M0,-0.25) node[below]{$|\Delta,-\Delta\rangle$};
\draw[->] (\XMIN,0) node[left]{\phantom{$\omega_{E}$}} -- (\XMAX,0) node[right]{$\omega_{E}$};
\draw[->] (0,-1) -- (0,0.75) node[above]{$\Delta(\Delta-1)$};
\draw (0,0) node[above left]{$0$};
\foreach \y in {-3,-2,-1,1,2,3} \draw (\y,0) node[above]{$\y$};
\draw (0,-0.25) node[below left]{$-\tfrac{1}{4}$};
\draw[dashed] (\XMIN,-0.25) -- (\XMAX,-0.25);
\filldraw[fill=white] (0,0) circle (0.8pt) node[above right]{$|0,0\rangle$};
\end{tikzpicture}
\label{fig:1c}}
\caption{Projective unitary representations of the Lie algebra $\mathfrak{sl}(2,\mathbb{R})_{A}$. Non-negativity of the squared norms $\|A_{\pm}|\Delta,\omega_{E}\rangle\|^{2} = \omega_{E}(\omega_{E} \pm 1) - \Delta(\Delta-1) \geq 0$ (see (a)) and the ladder equations $A_{\pm}|\Delta,\omega_{E}\rangle \propto |\Delta,\omega_{E}\pm1\rangle$ restrict the possible range of eigenvalues $\Delta(\Delta-1)$ and $\omega_{E}$ to the five distinct regions: (i) $C_{p}(\nu,\omega_{0})_{A}$ (light gray region in (b)), (ii) $C_{s}(\Delta, \omega_{0})_{A}$ (light blue region in (b)), (iii) $D_{+}(\Delta)_{A}$ (solid dark green curves in (c)), (iv) $D_{-}(\Delta)_{A}$ (solid dark red curves in (c)), and (v) trivial representation (white hole circle in (c)).}
\label{fig:1}
\end{figure}

Unitary representations of the Lie algebra $\mathfrak{sl}(2,\mathbb{R})_{A} \oplus \mathfrak{sl}(2,\mathbb{R})_{B}$ are constructed on the Hilbert space whose basis vectors are given by the tensor product $|\Delta, \omega_{E}\rangle \otimes |\Delta^{\prime}, \alpha_{E}\rangle$, where $|\Delta, \omega_{E}\rangle$ and $|\Delta^{\prime}, \alpha_{E}\rangle$ are basis of the Hilbert spaces for some unitary representations of $\mathfrak{sl}(2,\mathbb{R})_{A}$ and $\mathfrak{sl}(2,\mathbb{R})_{B}$, respectively.
Since the quadratic Casimirs of $\mathfrak{sl}(2,\mathbb{R})_{A}$ and $\mathfrak{sl}(2,\mathbb{R})_{B}$ coincide, the conformal weights $\Delta$ and $\Delta^{\prime}$ should also coincide, $\Delta = \Delta^{\prime}$.
Hence it is enough to study unitary representations of each $\mathfrak{sl}(2,\mathbb{R})$ separately.
Standard ``single-valued'' unitary representations of the Lie algebra $\mathfrak{sl}(2,\mathbb{R})$ are all classified by Bargmann \cite{Bargmann:1946me}.
Here ``single-valued'' means that an element of the Hilbert space satisfies the periodic boundary condition with respect to the compact dimension $S^{1} \cong SO(2)$.
As we have seen in section \ref{sec:2}, however, due to the presence of background electric field, the charged scalar field should satisfy the twisted boundary condition with respect to the imaginary time direction, $\Phi|_{t_{E} = 2\pi} = \mathrm{e}^{i2\pi\alpha_{E}}\Phi|_{t_{E} = 0}$ (tilde is omitted).
In this note we also impose the twisted boundary condition with respect to the fictitious compact dimension $S^{1} \ni \theta$, $\Phi|_{\theta = 2\pi} = \mathrm{e}^{-i2\pi\alpha_{0}}\Phi|_{\theta = 0}$, which physically describes the situation where a nonzero constant magnetic field ($\propto \alpha_{0}$) penetrates through the extra dimension $S^{1}$.
The presence of such constant magnetic field enables us to consider arbitrary value of Wick-rotated coupling constant $\alpha_{E}$.
(Without such constant magnetic field, $\alpha_{E}$ turns out to be integer multiples of $2\pi$.)
Relevant unitary representations to our problem are therefore ``multi-valued'' unitary representations, or \textit{projective} unitary representations, of the Lie algebra $\mathfrak{sl}(2,\mathbb{R})$, which are classified by Puk\'{a}nszky \cite{Pukanszky:1964} in the context of unitary representations for the universal covering group of $SL(2,\mathbb{R})$.
Let us first focus on $\mathfrak{sl}(2,\mathbb{R})_{A}$.
Apart from the trivial representation ($\Delta = \omega_{E} = 0$), there are four distinct projective unitary representations of the Lie algebra $\mathfrak{sl}(2,\mathbb{R})_{A}$ (see Figure \ref{fig:1} and its caption):
\begin{subequations}
\begin{enumerate}
\item \textit{Principal series representation $C_{p}(\nu, \omega_{0})_{A}$:} A two-parameter family of continuous series representations for $\Delta(\Delta - 1) < -\tfrac{1}{4}$.
The Hilbert space of this representation is spanned by the basis
\begin{align}
\bigl\{
|\Delta, \omega_{E}\rangle
\mid
\omega_{E} = \omega_{0} + n \quad (n \in \mathbb{Z})
\bigr\}, \label{eq:3.9a}
\end{align}
where $\Delta = \tfrac{1}{2} + i\nu$ with $\nu \in (0, \infty)$ and $\omega_{0} \in [-\tfrac{1}{2}, \tfrac{1}{2})$; see Figure \ref{fig:1b}.
In the coordinate realization \eqref{eq:3.3a} and \eqref{eq:3.3b} an element of the Hilbert space satisfies the boundary condition $\Phi|_{t_{E}=2\pi} = \mathrm{e}^{-i2\pi\omega_{0}}\Phi|_{t_{E}=0}$.
(Note that the eigenfunction of $A_{3} = i\partial_{t_{E}}$ is $\mathrm{e}^{-i\omega_{E}t_{E}}$.)

\item \textit{Supplementary series representation $C_{s}(\Delta, \omega_{0})_{A}$:} A two-parameter family of continuous series representations for $-\tfrac{1}{4} \leq \Delta(\Delta - 1) < 0$.
The Hilbert space of this representation is spanned by the basis
\begin{align}
\bigl\{
|\Delta, \omega_{E}\rangle
\mid
\omega_{E} = \omega_{0} + n \quad (n \in \mathbb{Z})
\bigr\}, \label{eq:3.9b}
\end{align}
where $\Delta \in (0, \tfrac{1}{2}]$ and $\omega_{0} \in (-\Delta, \Delta)$; see Figure \ref{fig:1b}.
In the coordinate realization \eqref{eq:3.3a} and \eqref{eq:3.3b} an element of the Hilbert space satisfies the boundary condition $\Phi|_{t_{E}=2\pi} = \mathrm{e}^{-i2\pi\omega_{0}}\Phi|_{t_{E}=0}$.

\item \textit{Positive discrete series representation $D_{+}(\Delta)_{A}$:} A one-parameter family of the lowest weight representations of lowest weight $\omega_{E} = \Delta$ for $\Delta(\Delta - 1) \geq -\tfrac{1}{4}$.
The Hilbert space of this representation is spanned by the basis
\begin{align}
\bigl\{
|\Delta, \omega_{E}\rangle
\mid
\omega_{E} = \Delta + n \quad (n \in \mathbb{Z}_{\geq 0})
\bigr\}, \label{eq:3.9c}
\end{align}
where $\Delta \in (0, \infty)$; see Figure \ref{fig:1c}.
In the coordinate realization \eqref{eq:3.3a} and \eqref{eq:3.3b} an element of the Hilbert space satisfies the boundary condition $\Phi|_{t_{E}=2\pi} = \mathrm{e}^{-i2\pi\Delta}\Phi|_{t_{E}=0}$.

\item \textit{Negative discrete series representation $D_{-}(\Delta)_{A}$:} A one-parameter family of the highest weight representations of highest weight $\omega_{E} = -\Delta$ for $\Delta(\Delta - 1) \geq -\tfrac{1}{4}$.
The Hilbert space of this representation is spanned by the basis
\begin{align}
\bigl\{
|\Delta, \omega_{E}\rangle
\mid
\omega_{E} = -\Delta - n \quad (n \in \mathbb{Z}_{\geq 0})
\bigr\}, \label{eq:3.9d}
\end{align}
where $\Delta \in (0, \infty)$; see Figure \ref{fig:1c}.
In the coordinate realization \eqref{eq:3.3a} and \eqref{eq:3.3b} an element of the Hilbert space satisfies the boundary condition $\Phi|_{t_{E}=2\pi} = \mathrm{e}^{i2\pi\Delta}\Phi|_{t_{E}=0}$.
\end{enumerate}
\end{subequations}
Similar results hold true for the projective unitary representations of $\mathfrak{sl}(2,\mathbb{R})_{B}$, which we denote by $C_{p}(\nu, \alpha_{0})_{B}$, $C_{s}(\Delta, \alpha_{0})_{B}$, $D_{+}(\Delta)_{B}$ and $D_{-}(\Delta)_{B}$, respectively.
Projective unitary representations of the Lie algebra $\mathfrak{sl}(2,\mathbb{R})_{A} \oplus \mathfrak{sl}(2,\mathbb{R})_{B}$ are then given by the tensor products of those for $\mathfrak{sl}(2,\mathbb{R})_{A}$ and $\mathfrak{sl}(2,\mathbb{R})_{B}$.
Notice that there are four possible combinations consistent with the range of $\Delta(\Delta-1)$ and the twisted boundary conditions $\Phi|_{t_{E}=2\pi} = \mathrm{e}^{i2\pi\alpha_{E}}\Phi|_{t_{E}=0}$ and $\Phi|_{\theta=2\pi} = \mathrm{e}^{-i2\pi\alpha_{0}}\Phi|_{\theta=0}$: the continuous series representation $C_{p}(\nu, -\alpha_{0})_{A} \otimes C_{p}(\nu, \alpha_{0})_{B}$, the supplementary series representation $C_{s}(\Delta, -\alpha_{0})_{A} \otimes C_{s}(\Delta, \alpha_{0})_{B}$, and the discrete series representations $D_{+}(\Delta)_{A} \otimes D_{-}(\Delta)_{B}$ and $D_{-}(\Delta)_{A} \otimes D_{+}(\Delta)_{B}$ with $\Delta = \alpha_{0} \in (0, \infty)$.

Now, which of these unitary representations are realized in our model?
To see this, we need to invoke the Schr\"odinger equation \eqref{eq:2.14}, which describes the negative-energy problem.
Let us, for a moment, assume that $\Delta(\Delta - 1) > 0$.
Then the Eckart potential $V(x) = \Delta(\Delta - 1)/\sinh^{2}x + 2\omega_{E}\alpha_{E}\coth x$ in \eqref{eq:2.14} is positive-definite unless $\omega_{E}\alpha_{E}$ is negative.
Obviously positive-definite potentials do not admit negative-energy states such that we must require $\omega_{E}\alpha_{E} < 0$.
There are only two possibilities for the solutions to the condition $\omega_{E}\alpha_{E} < 0$: i) $\omega_{E} > 0$ and $\alpha_{E} < 0$, or ii) $\omega_{E} < 0$ and $\alpha_{E} > 0$; that is, i) $D_{+}(\Delta)_{A} \otimes D_{-}(\Delta)_{B}$, or ii) $D_{-}(\Delta)_{A} \otimes D_{+}(\Delta)_{B}$.
In the unitary representation $D_{+}(\Delta)_{A} \otimes D_{-}(\Delta)_{B}$ the spectrum is given by
\begin{align}
\omega_{E} = \Delta + n
\quad\text{and}\quad
\alpha_{E} = - \Delta - n^{\prime}
\quad (n, n^{\prime} \in \mathbb{Z}_{\geq 0}), \label{eq:3.10}
\end{align}
while in the unitary representation $D_{-}(\Delta)_{A} \otimes D_{+}(\Delta)_{B}$ the spectrum is
\begin{align}
\omega_{E} = - \Delta - n
\quad\text{and}\quad
\alpha_{E} = \Delta + n^{\prime}
\quad (n, n^{\prime} \in \mathbb{Z}_{\geq 0}). \label{eq:3.11}
\end{align}
Notice that, since discrete series representations are valid for $\Delta(\Delta - 1) \geq -\tfrac{1}{4}$, the assumption $\Delta(\Delta - 1) > 0$ is in fact relaxed to the condition $\Delta(\Delta - 1) \geq -\tfrac{1}{4}$.\footnote{In this note we do not touch upon the case $\Delta(\Delta - 1) < -\tfrac{1}{4}$, where, just like the inverse square potential $V(x) = g/x^{2}$ with $g < -\tfrac{1}{4}$ \cite{Case:1950an}, the Eckart potential admits infinitely many negative-energy bound states (Efimov states).
As in the case of Efimov problem, when $\Delta(\Delta - 1) < -\tfrac{1}{4}$ the scale invariance is broken to the discrete scale invariance such that the spectrum could not be well-described by unitary representations of $\mathfrak{sl}(2,\mathbb{R})$.}

\section{Boundary spectrum generating algebra \texorpdfstring{$\mathfrak{sl}(2,\mathbb{R})_{A}^{0} \oplus \mathfrak{sl}(2,\mathbb{R})_{B}^{0}$}{sl(2,R) + sl(2,R)}} \label{sec:4}
In order to analyze near-boundary behaviors of the eigenstate $|\Delta, \alpha_{E}, \omega_{E}\rangle$, we need to know the asymptotic near-boundary behaviors of the operators \eqref{eq:3.3a}--\eqref{eq:3.4b}.
Noting that $\sinh x = x + O(x^{3})$ and $\coth x = 1/x + O(x)$ as $x \to 0$, and regarding the differential operator $\partial_{x}$ is of the order $O(1/x)$, we get the following boundary operators:
\begin{subequations}
\begin{align}
A_{3}^{0}
&:= 	\lim_{x\to0}A_{3}
= 	i\partial_{t_{E}}, \label{eq:4.1a}\\
A_{\pm}^{0}
&:= 	\lim_{x\to0}A_{\pm}
= 	\mathrm{e}^{\mp it_{E}}
	\left[
	\mp x\partial_{x} - i\partial_{t_{E}}
	\right], \label{eq:4.1b}\\
B_{3}^{0}
&:= 	\lim_{x\to0}B_{3}
= 	i\partial_{\theta}, \label{eq:4.1c}\\
B_{\pm}^{0}
&:= 	\lim_{x\to0}B_{\pm}
= 	\mathrm{e}^{\mp i\theta}
	\left[
	\mp x\partial_{x} - i\partial_{\theta}
	\right]. \label{eq:4.1d}
\end{align}
\end{subequations}
These operators still satisfy the commutation relations of the Lie algebra $\mathfrak{sl}(2,\mathbb{R}) \oplus \mathfrak{sl}(2,\mathbb{R}) \cong \mathfrak{so}(2,2)$
\begin{subequations}
\begin{alignat}{3}
&[A_{3}^{0}, A_{\pm}^{0}] = \pm A_{\pm}^{0},&\quad
&[A_{+}^{0}, A_{-}^{0}] = -2A_{3}^{0},& \label{eq:4.2a}\\
&[B_{3}^{0}, B_{\pm}^{0}] = \pm B_{\pm}^{0},&\quad
&[B_{+}^{0}, B_{-}^{0}] = -2B_{3}^{0},& \label{eq:4.2b}
\end{alignat}
\end{subequations}
with other commutators vanishing.
We denote this Lie algebra by $\mathfrak{sl}(2,\mathbb{R})_{A}^{0} \oplus \mathfrak{sl}(2,\mathbb{R})_{B}^{0}$.
The quadratic Casimir has the following simple form:
\begin{align}
C_{2}(\mathfrak{sl}(2,\mathbb{R})_{A}^{0})
= 	C_{2}(\mathfrak{sl}(2,\mathbb{R})_{B}^{0})
= 	x^{2}\partial_{x}^{2}
=: 	C_{2}^{0}. \label{eq:4.3}
\end{align}
Let $|\Delta, \omega_{E}, \alpha_{E}\rangle^{0}$ be a simultaneous eigenstate of $C_{2}^{0}$, $A_{3}^{0}$, and $B_{3}^{0}$ that satisfies the eigenvalue equations
\begin{subequations}
\begin{align}
C_{2}^{0}|\Delta,\omega_{E},\alpha_{E}\rangle^{0}
&= 	\Delta(\Delta-1)|\Delta,\omega_{E},\alpha_{E}\rangle^{0}, \label{eq:4.4a}\\
A_{3}^{0}|\Delta,\omega_{E},\alpha_{E}\rangle^{0}
&= 	\omega_{E}|\Delta,\omega_{E},\alpha_{E}\rangle^{0}, \label{eq:4.4b}\\
B_{3}^{0}|\Delta,\omega_{E},\alpha_{E}\rangle^{0}
&= 	\alpha_{E}|\Delta,\omega_{E},\alpha_{E}\rangle^{0}, \label{eq:4.4c}
\end{align}
\end{subequations}
and the ladder equations
\begin{subequations}
\begin{align}
A_{\pm}^{0}|\Delta, \omega_{E}, \alpha_{E}\rangle^{0}
&= 	\sqrt{(\omega_{E} \pm \Delta)(\omega_{E} \pm 1 \mp \Delta)}
	|\Delta, \omega_{E} \pm 1, \alpha_{E}\rangle^{0}, \label{eq:4.5a}\\
B_{\pm}^{0}|\Delta, \omega_{E}, \alpha_{E}\rangle^{0}
&= 	\sqrt{(\alpha_{E} \pm \Delta)(\alpha_{E} \pm 1 \mp \Delta)}
	|\Delta, \omega_{E}, \alpha_{E} \pm 1\rangle^{0}. \label{eq:4.5b}
\end{align}
\end{subequations}
In the coordinate realization \eqref{eq:4.1a}--\eqref{eq:4.1d} the eigenvalue equations \eqref{eq:4.4a}--\eqref{eq:4.4b} become the following differential equations:
\begin{subequations}
\begin{align}
\left(
-\partial_{x}^{2} + \frac{\Delta(\Delta-1)}{x^{2}}
\right)\Phi_{\Delta, \omega_{E}, \alpha_{E}}^{0}
&= 	0, \label{eq:4.6a}\\
\left(i\partial_{t_{E}} - \omega_{E}\right)\Phi_{\Delta, \omega_{E}, \alpha_{E}}^{0}
&= 	0, \label{eq:4.6b}\\
\left(i\partial_{\theta} - \alpha_{E}\right)\Phi_{\Delta, \omega_{E}, \alpha_{E}}^{0}
&= 	0, \label{eq:4.6c}
\end{align}
\end{subequations}
where $\Phi_{\Delta, \omega_{E}, \alpha_{E}}^{0}(x, t_{E}, \theta) = \langle x, t_{E}, \theta|\Delta, \omega_{E}, \alpha_{E}\rangle^{0}$.
These differential equations are easily solved with the result
\begin{align}
\Phi_{\Delta, \omega_{E}, \alpha_{E}}^{0}(x, t_{E}, \theta)
&= 	A_{\Delta}(\omega_{E}, \alpha_{E})
	x^{\Delta}\mathrm{e}^{-i\omega_{E}t_{E}}\mathrm{e}^{-i\alpha_{E}\theta}
	+ B_{\Delta}(\omega_{E}, \alpha_{E})
	x^{1-\Delta}\mathrm{e}^{-i\omega_{E}t_{E}}\mathrm{e}^{-i\alpha_{E}\theta}, \label{eq:4.7}
\end{align}
where the integration constants $A_{\Delta}(\omega_{E}, \alpha_{E})$ and $B_{\Delta}(\omega_{E}, \alpha_{E})$ may depend on $\Delta$, $\omega_{E}$ and $\alpha_{E}$.

\section{Recurrence relations for finite-temperature \texorpdfstring{CFT$_{1}$}{CFT1} two-point functions} \label{sec:5}
Now we are in a position to derive recurrence relations for Euclidean two-point functions in frequency space.
By applying the boundary ladder operators \eqref{eq:4.1b} and \eqref{eq:4.1d} to the solution \eqref{eq:4.7} we get
\begin{subequations}
\begin{align}
A_{\pm}^{0}\Phi_{\Delta, \omega_{E}, \alpha_{E}}^{0}
&= 	-(\omega_{E} \pm \Delta)A_{\Delta}(\omega_{E}, \alpha_{E})
	x^{\Delta}
	\mathrm{e}^{-i(\omega_{E} \pm 1)t_{E}}
	\mathrm{e}^{-i\alpha_{E}\theta} \nonumber\\
&\qquad
	- (\omega_{E} \pm 1 \mp \Delta)B_{\Delta}(\omega_{E}, \alpha_{E})
	x^{1-\Delta}
	\mathrm{e}^{-i(\omega_{E} \pm 1)t_{E}}
	\mathrm{e}^{-i\alpha_{E}\theta}, \label{eq:5.1a} \\
B_{\pm}^{0}\Phi_{\Delta, \omega_{E}, \alpha_{E}}^{0}
&= 	-(\alpha_{E} \pm \Delta)A_{\Delta}(\omega_{E}, \alpha_{E})
	x^{\Delta}
	\mathrm{e}^{-i\omega_{E}t_{E}}
	\mathrm{e}^{-i(\alpha_{E} \pm 1)\theta} \nonumber\\
&\qquad
	- (\alpha_{E} \pm 1 \mp \Delta)B_{\Delta}(\omega_{E}, \alpha_{E})
	x^{1-\Delta}
	\mathrm{e}^{-i\omega_{E}t_{E}}
	\mathrm{e}^{-i(\alpha_{E} \pm 1)\theta}. \label{eq:5.1b}
\end{align}
\end{subequations}
Comparing these to the right-hand side of the ladder equations \eqref{eq:4.5a} and \eqref{eq:4.5b} we get the following recurrence relations for the coefficients:
\begin{subequations}
\begin{align}
A_{\Delta}(\omega_{E}, \alpha_{E})
&= 	-\sqrt{\frac{\omega_{E} \pm 1 \mp \Delta}{\omega_{E} \pm \Delta}}
	A_{\Delta}(\omega_{E} \pm 1, \alpha_{E}), \label{eq:5.2a} \\
B_{\Delta}(\omega_{E}, \alpha_{E})
&= 	-\sqrt{\frac{\omega_{E} \pm \Delta}{\omega_{E} \pm 1 \mp \Delta}}
	B_{\Delta}(\omega_{E} \pm 1, \alpha_{E}), \label{eq:5.2b} \\
A_{\Delta}(\omega_{E}, \alpha_{E})
&= 	-\sqrt{\frac{\alpha_{E} \pm 1 \mp \Delta}{\alpha_{E} \pm \Delta}}
	A_{\Delta}(\omega_{E}, \alpha_{E} \pm 1), \label{eq:5.2c} \\
B_{\Delta}(\omega_{E}, \alpha_{E})
&= 	-\sqrt{\frac{\alpha_{E} \pm \Delta}{\alpha_{E} \pm 1 \mp \Delta}}
	B_{\Delta}(\omega_{E}, \alpha_{E} \pm 1). \label{eq:5.2d}
\end{align}
\end{subequations}
Euclidean two-point functions for charged scalar operator of conformal weight $\Delta$ in the dual CFT$_{1}$, which is given by the ratio $G^{E}_{\Delta}(\omega_{E}, \alpha_{E}) = (2\Delta - 1)A_{\Delta}(\omega_{E}, \alpha_{E})/B_{\Delta}(\omega_{E}, \alpha_{E})$, then satisfy the following recurrence relations:
\begin{subequations}
\begin{align}
G^{E}_{\Delta}(\omega_{E}, \alpha_{E})
&= 	\frac{\omega_{E} \pm 1 \mp \Delta}{\omega_{E} \pm \Delta}
	G^{E}_{\Delta}(\omega_{E} \pm 1, \alpha_{E}), \label{eq:5.3a} \\
G^{E}_{\Delta}(\omega_{E}, \alpha_{E})
&= 	\frac{\alpha_{E} \pm 1 \mp \Delta}{\alpha_{E} \pm \Delta}
	G^{E}_{\Delta}(\omega_{E}, \alpha_{E} \pm 1), \label{eq:5.3b}
\end{align}
\end{subequations}
which are easily solved by iteration.
Let us first focus on the positive-frequency solutions $\omega_{E} = \Delta + n$ and $\alpha_{E} = - \Delta - n^{\prime}$ ($n, n^{\prime} \in \mathbb{Z}_{\geq 0}$).
The iterative use of the relations $G^{E}_{\Delta}(\omega_{E}, \alpha_{E}) = \frac{\Delta - 1 + \omega_{E}}{-\Delta + \omega_{E}}G^{E}_{\Delta}(\omega_{E} - 1, \alpha_{E})$ and $G^{E}_{\Delta}(\omega_{E}, \alpha_{E}) = \frac{\Delta - 1 - \alpha_{E}}{-\Delta - \alpha_{E}}G^{E}_{\Delta}(\omega_{E}, \alpha_{E} + 1)$ gives
\begin{align}
G^{E}_{\Delta}(\omega_{E}, \alpha_{E})
&= 	\frac{\Gamma(\Delta + \omega_{E})}{\Gamma(1 - \Delta + \omega_{E})}
	\frac{\Gamma(\Delta - \alpha_{E})}{\Gamma(1 - \Delta - \alpha_{E})}
	\frac{1}{\Gamma(2\Delta)^{2}}G^{E}_{\Delta}(\Delta, -\Delta). \label{eq:5.4}
\end{align}
Restoring $r_{0}$ via $\omega_{E} \to \omega_{E}r_{0}$ and analytically continuing back to Lorentzian signature by $\omega_{E}r_{0} = -i(\omega r_{0} + \alpha)$ and $\alpha_{E} = -i\alpha$, we get the retarded two-point function
\begin{align}
G^{R}_{\Delta}(\omega, \alpha)
= 	\frac{\Gamma(\Delta - \frac{i\omega}{2\pi T} - i\alpha)}
	{\Gamma(1 - \Delta - \frac{i\omega}{2\pi T} - i\alpha)}
	\frac{\Gamma(\Delta + i\alpha)}{\Gamma(1 - \Delta + i\alpha)}
	g^{R}(\Delta), \label{eq:5.5}
\end{align}
where $g^{R}(\Delta)$ is a both $\omega$- and $\alpha$-independent normalization factor which should be determined from $G^{E}_{\Delta}(\Delta, -\Delta)/\Gamma(2\Delta)^{2}$ and $T$ is the Hawking temperature given by
\begin{align}
T
&= 	\frac{1}{2\pi r_{0}}. \label{eq:5.6}
\end{align}
Notice that Eq.~\eqref{eq:5.5} indeed has a desired analytical structure of retarded two-point function; that is, it is analytic in the upper-half complex $\omega$-plane and has simple poles at $\omega = - 2\pi T\alpha - i2\pi T(\Delta + n)$ ($n \in \mathbb{Z}_{\geq 0}$) in the lower-half complex $\omega$-plane.
As generally discussed in \cite{Son:2002sd}, these simple poles correspond to the quasi-normal frequencies for charged scalar fields on AdS$_{2}$ black hole.
Note also that, up to the overall normalization factor $g^{R}(\Delta)$, Eq.~\eqref{eq:5.5} precisely coincides with the result obtained by Faulkner \textit{et al.} \cite{Faulkner:2009wj,Faulkner:2011tm}.\footnote{\label{footnote:6}Precisely speaking, there is a sign difference in $\alpha$ between our retarded/advanced two-point functions and the results in \cite{Faulkner:2009wj,Faulkner:2011tm}.
This difference can be understood as follows.
In \cite{Faulkner:2009wj,Faulkner:2011tm} the authors considered the Reissner-Nordstr\"{o}m-AdS black hole, whose near-horizon geometry has the following AdS$_{2}$ factor:
\begin{align}
ds_{\text{AdS}_{2}}^{2}
= 	\frac{R^{2}}{r^{2}}
	\left[
	-\left(1 - \left(\frac{r}{r_{0}}\right)^{2}\right)dt^{2}
	+ \frac{dr^{2}}{1 - (r/r_{0})^{2}}
	\right]
\quad\text{with}\quad
A
= 	\frac{ER^{2}}{r}\left(1 - \frac{r}{r_{0}}\right)dt, \nonumber
\end{align}
where $R$ is the AdS$_{2}$ radius, $r_{0}$ is the horizon radius and $E$ is the constant electric field.
(In the notation of \cite{Faulkner:2009wj,Faulkner:2011tm} $ER^{2} = e_{d}$.)
By changing the spatial coordinate via $r = r_{0}\tanh(x/R)$ and further rescaling the time as $t = (r_{0}/R)\Tilde{t}$, the metric becomes the conformal metric
\begin{align}
ds_{\text{AdS}_{2}}^{2}
= 	\frac{-d\Tilde{t}^{2} + dx^{2}}{\sinh^{2}(x/R)}
\quad\text{with}\quad
A
= 	ER(\coth(x/R) - 1)d\Tilde{t}. \nonumber
\end{align}
Notice the sign difference in the gauge field $A$ between \eqref{eq:1.3} and the above expression.
By replacing $\alpha$'s in \eqref{eq:5.5} and \eqref{eq:5.7} with $-\alpha = -qER^{2}$ we get the results in \cite{Faulkner:2009wj,Faulkner:2011tm}.
(Note that the frequency $\omega$ conjugate to the original time $t$ and the frequency $\Tilde{\omega}$ conjugate to the rescaled time $\Tilde{t}$ are related by $\Tilde{\omega} = (r_{0}/R)\omega$.
Hence we have $\omega/2\pi T = \Tilde{\omega}/2\pi\Tilde{T}$, where $T = 1/2\pi r_{0}$ and $\Tilde{T} = 1/2\pi R$ are Hawking temperatures with respect to the coordinates $t$ and $\Tilde{t}$, respectively.)}
The advanced two-point function is similarly obtained by starting from the negative-frequency solutions $\omega_{E} = -\Delta - n$ and $\alpha_{E} = \Delta + n^{\prime}$ ($n, n^{\prime} \in \mathbb{Z}_{\geq 0}$).
The result is
\begin{align}
G^{A}_{\Delta}(\omega, \alpha)
= 	\frac{\Gamma(\Delta + \frac{i\omega}{2\pi T} + i\alpha)}
	{\Gamma(1 - \Delta + \frac{i\omega}{2\pi T} + i\alpha)}
	\frac{\Gamma(\Delta - i\alpha)}{\Gamma(1 - \Delta - i\alpha)}
	g^{A}(\Delta), \label{eq:5.7}
\end{align}
where $g^{A}(\Delta)$ is an overall normalization factor, which should be related to $g^{R}(\Delta)$ by complex conjugate $g^{A}(\Delta) = [g^{R}(\Delta)]^{\ast}$ since retarded and advanced two-point functions are in general related by $G^{A}_{\Delta}(\omega, \alpha) = [G^{R}_{\Delta}(\omega, \alpha)]^{\ast}$.

\section{Conclusions and discussions} \label{sec:6}
Much inspired by algebraic approach to S-matrix in quantum mechanics \cite{Frank:1984,Alhassid:1984uy}, in this note we have presented a new Lie-algebraic method to compute finite-temperature CFT$_{1}$ two-point functions in frequency space by making use of the real-time prescription of AdS/CFT correspondence \cite{Iqbal:2008by,Iqbal:2009fd}.
We have shown that frequency- and charge-dependences of two-point functions for charged scalar operators of CFT$_{1}$ dual to AdS$_{2}$ black hole with constant background electric field is completely determined by Lie-algebraic manipulations.
In contrast to the standard AdS/CFT techniques, our algebraic method does not require to solve the bulk field equation explicitly: What we have to know are unitary representations realized in the model, asymptotic near-boundary algebra $\mathfrak{sl}(2,\mathbb{R})_{A}^{0} \oplus \mathfrak{sl}(2,\mathbb{R})_{B}^{0}$ and its coordinate realization \eqref{eq:4.1a}--\eqref{eq:4.1d} given in section \ref{sec:4}.

Let us close with comments on some possible future directions in Lie-algebraic approach to momentum-space CFT two-point functions.
Though in this note we have focused on a scalar field theory for simplicity, it is interesting to generalize our method to two-point functions for fermion fields, vector fields and tensor fields.
This should be possible by using transformation laws of these fields under the action of isometry group of AdS spacetime, because coordinate realizations of infinitesimal generators are enough to obtain asymptotic near-boundary algebra.
It is also interesting to generalize our method to AdS$_{d+1}$/CFT$_{d}$ correspondence with $d > 1$.
For example, it is easy to generalize our method to three-dimensional BTZ black hole, which is locally AdS$_{3}$ and has $SL(2,\mathbb{R})_{L} \times SL(2,\mathbb{R})_{R}$ symmetry, and compute momentum-space two-point functions of dual CFT$_{2}$ at finite-temperature by just using a differential operator realization of the Lie algebra $\mathfrak{sl}(2,\mathbb{R})_{L} \oplus \mathfrak{sl}(2,\mathbb{R})_{R} (\cong \mathfrak{so}(2,2))$ and its unitary representations.\footnote{Computational details will be presented in a future publication.}
Such momentum-space two-point functions are already computed in \cite{Birmingham:2001pj,Iqbal:2009fd,Balasubramanian:2010sc} by explicitly Fourier-transforming the position-space two-point functions or by explicitly solving the bulk field equations on the BTZ black hole background.
The resultant momentum-space two-point functions are known to have nice factorized forms for left- and right-moving sectors, and indeed have essentially the same forms as those for charged scalar operators of CFT$_{1}$ governed by $SL(2,\mathbb{R})_{A} \times SL(2,\mathbb{R})_{B}$ symmetry.

Finally, most of our method has a common structure to Lie-algebraic approach to S-matrix with dynamical symmetry, however, there still exists a gap between them.
Though our method is based on a specific coordinate realization of the Lie algebra, it is known that algebraic approach to S-matrix is free from specific coordinate realizations and can be formulated in much more algebraic ways.
As far as we know, there are two distinct such purely algebraic approaches to scattering problems.
One is a method called the Euclidean connection developed by Alhassid and his collaborators \cite{Alhassid:1984uy,Alhassid:1984,Alhassid:1986,Frank:1986}, which is based on group contractions and expansions, and the other is a method developed by Kerimov \cite{Kerimov:1998zz} (see also \cite{Kerimov:1998,Kerimov:2002}), which is based on intertwining operators between Weyl equivalent principal series representations of dynamical groups.
Since it is already known in the context of both CFT \cite{Koller:1974ut} and AdS/CFT correspondence \cite{Dobrev:1998md} that two-point functions can be regarded as kernels of intertwining operators between two representations of conformal group conjugate with each other by Weyl reflection, it is natural to expect that one can also formulate purely algebraic methods to compute momentum-space CFT two-point functions along the line of Kerimov's method.
We would like to leave this issue for future studies.

\subsection*{Acknowledgements}
The author would like to thank Masahide Manabe for conversations.
This work is supported in part by EU project CZ.1.07/2.3.00/30.0034.

\bibliographystyle{utphys}
\bibliography{Bibliography}
\end{document}